\documentclass{myaipproc}

\newcommand{\lsim}{\mathrel{<\kern-1.0em\lower0.9ex\hbox{$\sim$}}}
\newcommand{\gsim}{\mathrel{>\kern-1.0em\lower0.9ex\hbox{$\sim$}}}

\def\paperversion{}			

\begin{document}

\title{Large-Scale Simulations of Clusters of Galaxies}
\author{P.~M.~Ricker\affiliationmark{a},
A.~C.~Calder\affiliationmark{a},
L.~J.~Dursi\affiliationmark{a},
B.~Fryxell\affiliationmark{a},
D.~Q.~Lamb\affiliationmark{a},
P.~MacNeice\affiliationmark{b},
K.~Olson\affiliationmark{a,b},
R.~Rosner\affiliationmark{a},
F.~X.~Timmes\affiliationmark{a},
J.~W.~Truran\affiliationmark{a},
H.~M.~Tufo\affiliationmark{a}, and
M.~Zingale\affiliationmark{a}
}
\affiliation{
  \affiliationmark{a}ASCI Flash Center, University of Chicago,
	Chicago, IL 60637\\
  \affiliationmark{b}NASA/Goddard Space Flight Center, Greenbelt, MD 20771
}

\begin{abstract}
We discuss some of the computational challenges encountered in simulating
the evolution of clusters of galaxies.
Eulerian adaptive mesh refinement (AMR) techniques can successfully
address these challenges but are currently being used by only a few groups.
We describe our publicly available AMR code, FLASH, which
uses an object-oriented framework to manage its AMR library,
physics modules, and automated verification.
We outline the development of the FLASH framework to
include collisionless particles, permitting it to be used for cluster simulation.
\end{abstract}

\maketitle


\section{Simulating Clusters}

Clusters of galaxies are the largest gravitationally bound objects in
the universe.
They consist mainly of dark matter and diffuse, hot plasma, with
galaxies themselves contributing only a few percent of the total mass.
Clusters have attracted attention in recent years
because they are large enough to serve
as a representative sample of the universe; they provide strong contraints
on cosmological models.
Clusters are also interesting from an astrophysical point of view.
The intracluster medium (ICM), at densities
$\sim 10^{-4}-10^{-2}$~cm$^{-3}$ and temperatures $\sim 10^{7-8}$~K,
is collisionally ionized and emits X-rays, primarily via
bremsstrahlung \cite{Sar88}.
The radiative cooling time can be short enough to produce cooling flows \cite{Fab94}.
Observations of Faraday rotation show that the ICM is magnetized,
with $B\gsim 1\ \mu$G \cite{Cla00}.
With the diffuse, nonthermal radio \cite{KemSar00}
and X-ray \cite{FF99} emission seen in some clusters,
this suggests that clusters are sites for cosmic-ray acceleration \cite{BBP97}.
Magnetic fields may also help to suppress diffusion
in the ICM \cite{ChaCow99}.
Galaxies orbiting in cluster potentials experience tidal and ram-pressure
stripping and help to stir the ICM \cite{SAP99}.
Star formation and supernovae may affect the abundances of heavy elements
in the ICM as well as its global energetics \cite{DupWhi00,LPC00}.
Many elements of a complete model for the ICM are known,
but we still cannot answer such questions as:
what is the source of entropy and metals in the ICM?
what happens to the gas that cools below X-ray temperatures in cooling flows?
how robust are cooling flows?
how is energy partitioned among thermal and nonthermal particle
populations, magnetic fields, and turbulent motions?

Cluster mergers play a key role in these phenomena.
Mergers strongly affect the ICM, producing
long-lived distortions in X-ray images and temperature maps.
The energies they release ($\sim 10^{64}$~erg) are easily enough to heat
the ICM to $10^8$~K.
These events are complex, nonlinear, multiphysical, and three-dimensional;
thus numerical simulations are appropriate tools for studying them.

\begin{figure}
\includegraphics{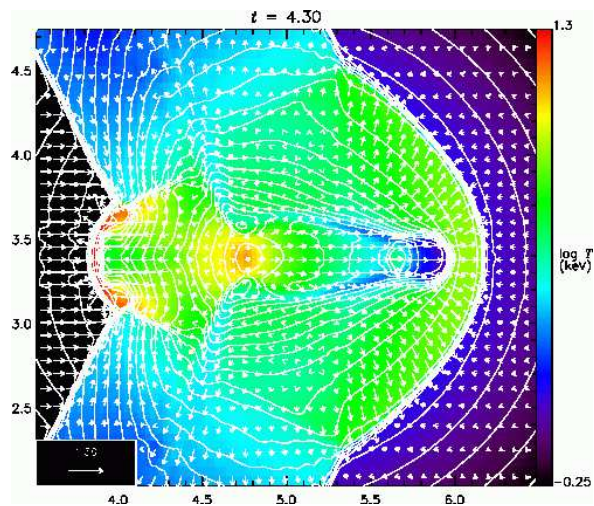}
\caption{Gas density (log contours), temperature (shading), and velocity
(arrows) in a
merger of clusters with mass ratio 1:3 (Ricker \& Sarazin 2000,
in prep.). Times and lengths are in Gyr and $h^{-1}$~Mpc,
respectively, with Hubble constant $h=0.6$.}
\label{Fig:merger}
\end{figure}

Cluster simulations with multiphysics (at least hydrodynamics in addition
to dark matter) have reached $\gsim 10^6$ particles or
zones, simulating a cluster and its
environment down to kpc scales \cite{Fre99,Lew00}.
Cosmological simulations with $10^9$ particles,
yielding $10^5$ clusters, have been
performed \cite{Col00}, but thus far these have only included
dark matter.
To understand recent observations, we must resolve scales needed
for cosmological context ($\gsim$10~Mpc) and galaxies ($\lsim$1~kpc)
-- a dynamic range of $>10^4$ --
with hydrodynamics,
cooling, magnetic fields, and nonequilibrium plasma effects.
Star formation and supernova feedback will remain unresolved
for the forseeable future and must be included phenomenologically.

Because shocks are important in the ICM, Eulerian shock-capturing
methods such as the Piecewise-Parabolic Method (PPM) \cite{ColWoo84} are very
desirable.
Fig.~\ref{Fig:merger} shows an example merger calculation performed
using the COSMOS $N$-body/hydro code \cite{RDL00}. COSMOS uses PPM on a
single nonuniform grid.
This 3D calculation covered a dynamic range of $\sim 300$ on 128 processors
of the San Diego Cray T3E, requiring $10,000$ node-hours.
To add new physics and increase dynamic range,
the cost of such calculations must be reduced significantly
while retaining the shock-capturing properties of single-grid Eulerian schemes.


\section{Adaptive Mesh Refinement}

The computational
issues involved in studying gravitational clustering with multiphysics
involve the coupling of small and large scales through gravity and hydrodynamics,
requiring large dynamic range, and the presence of short-range source terms that
upset load balancing.

\begin{figure}
\includegraphics{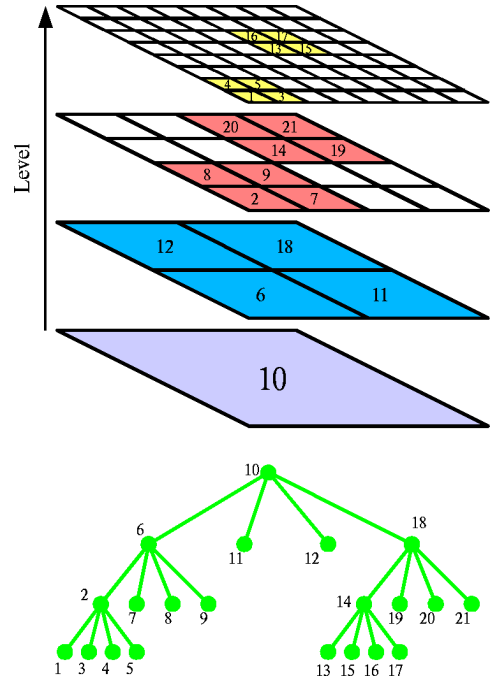}
\caption{Example 2D mesh managed by PARAMESH \protect\cite{Mac00}.}
\label{Fig:paramesh}
\end{figure}

Block-structured adaptive mesh refinement (AMR) methods address these issues
by placing fine grids only where they are needed to resolve fine features
\cite{BerOli84}.
An example of a freely available AMR package is PARAMESH \cite{Mac00}.
PARAMESH manages an octree (in 3D) data structure whose nodes are uniformly
gridded meshes (`blocks'); Fig.~\ref{Fig:paramesh} shows an example.
Each block is a factor of two more refined than its parent.
Refined blocks are placed according to user-defined criteria; interpolation is
used to obtain their initial and boundary data from
coarser blocks.
PARAMESH distributes blocks among processors using a work-weighted space-filling
curve, keeping spatially adjacent blocks on the same processor when possible
and balancing the computational load.

Long-range coupling can be handled in AMR using 
multilevel relaxation techniques \cite{BHM00}.
The coarse-grid solution is obtained on one processor, while
finer levels use neighbor-to-neighbor communication.

Short-range forces produce highly clustered distributions of work.
The mixed material representations in PPM-based cluster simulations
(Lagrangian particles, Eulerian gas) require different domain decompositions.
AMR can solve both of these problems by weighting blocks
appropriately, e.g., by source terms or particle content.
Blocks also can be evolved on different timesteps and weighted inversely
by their timestep.

AMR techniques are widely used in cosmological
$N$-body and smoothed-particle hydrodynamics
codes, but few groups have used them with Eulerian schemes \cite{NorBry99}.
AMR codes are difficult to construct,
and most cosmological codes are proprietary.
However, during the coming year we expect to see several AMR
codes useful for cosmology emerge, some of which are freely available.


\section{The FLASH Code}

We are developing FLASH, an adaptive-mesh astrophysical
simulation code based on PARAMESH \cite{Ros00,Fry00}.
FLASH is coded mainly in Fortran 90 and uses the Message-Passing Interface
(MPI).
It is highly portable and scales to thousands of processors.
We have recently been awarded the Gordon Bell Prize for achieving 0.24~TFlops
with FLASH
using 6,420 processors of ASCI Red on a cellular detonation problem relevant
to Type Ia supernovae \cite{Cal00}.
We intend for FLASH to evolve into a community
simulation framework; the code is publicly available at
{\tt http://flash.uchicago.edu/}.

Many astrophysical problems require
multiple physical processes and a wide range of scales.
Each physical process requires a different numerical method and different tests.
Exploiting AMR also requires complicated mesh management libraries.
Such complex software is best managed using a framework.

Object-oriented languages provide several features useful for
building simulation frameworks.  Encapsulation allows us to
interchange solvers that need conflicting internal data
structures; inheritance allows us to abstract common features of
different types of solvers; and polymorphism
allows us to switch between solvers.

Component frameworks scale better with
increasing complexity by providing standard
ways for components to describe themselves to each other.
Such frameworks are commonly used in business,
but they have not seen wide use in science, because they
impose unacceptable overhead and
lack features needed for scientific applications.  An
appropriate scientific component standard, such as that being
developed by the Common Component Architecture (CCA) Forum \cite{Arm99},
is still several years away.

\begin{figure}
\includegraphics{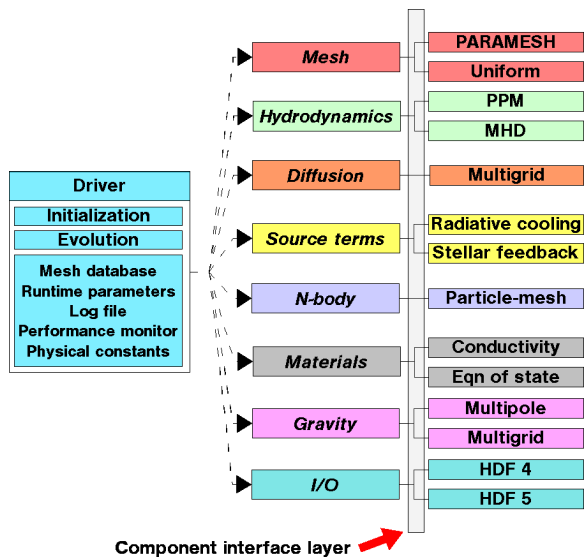}
\caption{The framework of the FLASH code \protect\cite{Fry00}, showing
components useful for cluster simulation.}
\label{Fig:flash}
\end{figure}

The FLASH framework is object-oriented and makes use of some component ideas.
Its class structure appears in Fig.~\ref{Fig:flash}.
The driver maintains mesh data in a static
container class and instantiates objects from various classes
of physics solvers.
The solvers are divided into different classes 
by their level of coupling and by differences in solution method
(e.g., hyperbolic solvers for
hydrodynamics, elliptic solvers for radiation and gravity).
The AMR library is also treated as a class.
Solver and mesh objects access
mesh data through methods supplied by the mesh
container class.
The component interface layer, for which we are developing
a standard, will consist of F90 module wrappers implementing
an interface that is
abstractly specified in an interface definition language (IDL).

FLASH includes hydrodynamics using PPM, self-gravity using
multigrid and multipole methods, and modules appropriate
for supernova problems, including a partially degenerate equation of state
and nuclear reaction networks. Modules for front tracking,
implicit diffusion, magnetohydrodynamics, and collisionless particles
are under active development by our group.
With these new components FLASH will be capable of simulating individual
clusters with multiphysics and a dynamic range of $\gsim 2,000$ per dimension
during the coming year.


\bigskip
\medskip

This work was supported by DOE under Grant No.\ B341495 to the ASCI
Flash Center at the University of Chicago.
Calculations were performed using the resources of the San Diego Supercomputer
Center.


\end{document}